\documentclass[10pt, conference]{IEEEtran}

\usepackage{amsmath}
\usepackage{amsfonts}
\usepackage{amssymb}
\usepackage{graphicx}
\usepackage{epsfig}
\usepackage[scriptsize,center]{caption}
\usepackage{subfig}
\usepackage{float}
\usepackage{epstopdf}
\usepackage{array}
\usepackage{multirow}
\usepackage{setspace}
\usepackage[section]{placeins}
\usepackage{url}
\usepackage{bbm}
\renewcommand\IEEEkeywordsname{Keywords}
\usepackage{cite}

\title{Listen Before Receive for Coexistence 
	in Unlicensed mmWave Bands}

\author{\IEEEauthorblockN{ 
		Sandra Lagen and Lorenza Giupponi
	}
	\IEEEauthorblockA{Mobile Networks department, Centre Tecnol\`ogic de Telecomunicacions de Catalunya, Castelldefels, Spain \\ e-mails: \{sandra.lagen, lorenza.giupponi\}@cttc.es}}

\begin{document}

\maketitle
                       
\begin{abstract}
Listen-Before-Talk (LBT) has been adopted as the spectrum sharing technique that guarantees a fair LTE/Wi-Fi coexistence in the unlicensed spectrum at the 5 GHz band. Differently, at mmWave bands, where beamforming is a must to overcome propagation limits, LBT scope becomes limited because the interference layout changes due to the directionality of transmissions. In this regard, this paper proposes a Listen-Before-Receive (LBR) technique for shared spectrum access and analyzes its potentials to promote a fair coexistence of multiple Radio Access Technologies (RATs) in unlicensed mmWave bands, as, e.g., 5G New Radio (NR) access technology and Wireless Gigabit (WiGig) devices using IEEE 802.11ad/ay standard. Since the less likely but still harmful interference situations with directional transmissions can no longer be detected easily at the transmitter, we believe that the receiver has useful information to be used. The main idea of LBR is that we provide to the receiver a say when it comes to allowing/preventing the access to the channel. In this line, we propose potential implementations of LBR, in conjunction with LBT and the self-contained slot, for NR-based access to unlicensed mmWave bands. 
\end{abstract}


\IEEEpeerreviewmaketitle
\section{Introduction}
\label{sec:intro}
To address the rapid and exponential increase of wireless data traffic demand in the next years, a recent trend in cellular networks is to consider both licensed and unlicensed bands to aggregate different portions of the spectrum, and thus improve the system capacity \cite{peng:14}. This was the case of LTE Licensed-Assisted Access (LTE-LAA) with the 5 GHz unlicensed band \cite{TR36889}. Notably, there has been a recent release of unlicensed spectrum at the 60 GHz band, which provides 10x times (in Europe) and 16x times (in the US) as much unlicensed spectrum bandwidth as available in sub 6 GHz bands \cite{5Gamericas}. Such spectrum will be exploited by the New Radio (NR) access technology that is being developed by 3GPP 5G standard \cite{TR38912}. NR-based access to unlicensed spectrum will be studied in NR Phase 2 through a recently approved study item \cite{RP-170828}. 

Despite the large bandwidth available at the 60 GHz millimeter wave (mmWave) band, the most critical issue of allowing cellular networks to operate in unlicensed spectrum is to ensure a fair and harmonious coexistence with the unlicensed systems, such as widely deployed Wi-Fi (IEEE 802.11a/h/j/n/ac) in the 5 GHz band and WiGig devices (IEEE 802.11ad/ay) in the 60 GHz band \cite{wigig}. In this regard, the Listen-Before-Talk (LBT) technique was proposed in LTE-LAA to ensure a fair coexistence with Wi-Fi \cite{TR36889}. LBT was designed assuming an omnidirectional transmit/receive pattern, and its key feature is that transmitters have to sense the medium before transmission, and they only transmit if the medium is sensed as idle, which is known as Clear Channel Assessment (CCA) in the IEEE 802.11 context. Besides, carrier sense (e.g., LBT) is a regulatory requirement in some regions like Europe and Japan for the 5 GHz and the 60 GHz bands \cite{TR38805,ETSI302567}, and has been adopted by standards like LTE-LAA \cite{TR36889}, IEEE 802.11 \cite{wigig}, and technologies like MuLTEFire \cite{multefire}.

\begin{figure}[!t]
	\centering
	\includegraphics[width=0.24\textwidth]{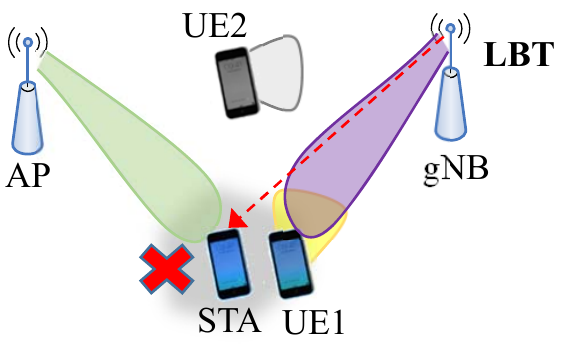}
	\caption{Interference problems in unlicensed mmWave bands with directional transmissions due to incorrect LBT. }
	\label{fig_hidnode}
	\vspace{-0.2cm}
\end{figure}

The major difference of NR/WiGig coexistence in the 60 GHz band with respect to LTE/Wi-Fi coexistence relies on the mmWave propagation characteristics, which impose the use of beamforming and directional transmissions to overcome propagation limits \cite{andrews:17}. 
The use of narrow beams enhances the spatial reuse and stimulates coexistence of different RATs, but it also changes the interference layout. In particular, the physical carrier sense of LBT can be performed either omnidirectionally (omniLBT) or directionally with the transmit (Tx) beam (dirLBT). 
OmniLBT is overprotective and depresses spatial reuse, because 
a transmission is prevented even if a signal is detected from a direction that may not create harmful interference for the intended receiver, while dirLBT enables spatial reuse but may create some hidden node problems. 

Even though, there are situations in which still on-going nearby transmissions are not detected at the transmitter, either with omniLBT or dirLBT, and hidden node problems that cause interference arise. For example, in Fig. \ref{fig_hidnode}, WiGig Access Point (AP) is transmitting towards a WiGig Station (STA) with its Tx beam (green beam). Then, NR Base Station (BS), a.k.a. gNB, wants to access the channel by performing LBT (dirLBT or omniLBT), which senses the channel as idle, and so gNB proceeds with directional data transmission towards the NR user (UE1) (purple beam). This way, gNB's transmission generates interference onto the STA (red arrow). Note that, if UE2 had been scheduled, no interference would have arised.

Research on NR/WiGig coexistence in the 60 GHz band has just started, and it is immature due to the on-going NR definition for 5G systems. Authors in \cite{nekovee:16} propose a solution that is based on iteratively coordinating the concurrent transmissions of different BSs by means of beam scheduling. 
In \cite[Sect. 8.2.2]{D41mmMAGIC}, a Listen-After-Talk (LAT) technique is introduced, in which the default mode for a transmitter is to send data and collisions detected by the receiver are solved according to coordination signaling. However, LAT is not compliant with the LBT requirement.
Wi-Fi and WiGig use an optional Ready to Send and Clear to Send (RTS/CTS) protocol to reduce intra-RAT collisions, but it requires virtual carrier sense and IEEE 802.11 RTS/CTS messages are not decodable by NR.
In the area of IEEE 802.15 Wireless Personal Area Networks (WPANs), multiple solutions have been proposed for beam management and time-domain coordination in mmWave bands with directional transmissions/receptions \cite{an:08,pyo:09,cai:10,singh:10}. 
Although either the coordination of the Tx beams (as proposed in \cite{nekovee:16,cai:10}) or the coordination of the channel access in time domain (as analyzed in \cite{an:08,pyo:09}) could solve hidden node problems, these kind of solutions require WiGig and NR coordination, which is not possible due to the asynchronous and autonomous operational mode of WiGig. For that reason, LBT was adopted to control channel accesses in LTE-LAA. However, omniLBT and dirLBT might be incorrect in the unlicensed mmWave bands because the interference dynamics are very different at the transmitter and receiver sides, as illustrated in Fig. \ref{fig_hidnode}. Therefore, new distributed and uncoordinated channel access schemes are needed to address NR/WiGig coexistence.

A key observation we make is that interference situations under directional transmissions/receptions might be more easily detected at the receiver, rather than at the transmitter, as it was assumed for LBT in the 5 GHz band. In this line, this paper presents a novel distributed technique, coined Listen-Before-Receive (LBR), to complement LBT, and in which the receiver has a say about whether it is appropriate or not for the transmitter to access the channel. In this way, we aim to guarantee a fair spectrum access for multi-RAT coexistence in unlicensed mmWave bands. LBR can be applied: \textit{i}) to a general multi-RAT environment, e.g., composed of WiGig, NR from operator A, and NR from operator B, and \textit{ii})  to manage the access of different uncoordinated BSs of the same RAT.

\section{System Model}
\label{sec:sysmodel}
Consider that 5G NR gNBs and WiGig APs coexist 
in the 60 GHz unlicensed mmWave band and that all of them use directional transmissions to overcome propagation limits. Each gNB intends to communicate to a NR UE, and every AP does so towards a WiGig STA. Except when needed, we do not differentiate among gNBs and APs, and refer to them as general BSs. Similarly, we refer to both UEs and STAs generically as Mobile Terminals (MTs).
Based on that, we consider a network deployment composed of $K$ BS-MT pairs that attempt access to the unlicensed spectrum.
Without loss of generality, we focus on the downlink (DL) transmission (transmission from BS to MT), as similar interference scenarios can be thought of also for uplink (UL) transmission or mixed DL-UL scenarios. Assume that every BS has $M$ transmit antennas, and each MT has $N$ receive antenna elements. 

We assume that beam-steering has been performed during a well-established beam-training phase, so that every $j$th BS has a Tx beam aligned towards its MT, which is denoted by $\textbf{w}_{j} {\in} \mathbb{C}^{M\times 1}$. For data decoding, every $k$th MT employs a receive (Rx) beam $\textbf{r}_{k} {\in} \mathbb{C}^{N\times 1}$. In IEEE 802.11ad, STAs might receive either omnidirectionally or directionally \cite{wigig}, so we assume that MTs might receive either with an omnidirectional or directional Rx beam, which is provided by the MT capability.

According to the standards, both in Wi-Fi and WiGig, CCA check is done omnidirectionally. eNBs/UEs in LTE follow LBT based on omnidirectional Energy Detection (ED), i.e., the so-called LBT Category 4 \cite{TR36889}, which resembles CCA in IEEE 802.11. LBT for NR has still not been defined. In this paper, we consider that all BSs perform CCA based on ED, and they can do so either omnidirectionally (omniLBT) or directionally with the Tx beam (dirLBT). In any case, due to the directional transmissions, omniLBT and dirLBT might be incorrect and BS's transmission might interfere to a nearby MT, as shown in Fig. \ref{fig_hidnode}.

The achievable data rate for the $k$th BS-MT link (in bits/s) is expressed as:
\vspace{-0.1cm}
\begin{equation}
R_k=W\log_2\Big(1+\frac{P_{k,k}}{N_oW{+}I_k}\Big) , \ I_k=\sum\nolimits_{j\ne k} P_{k,j}, \label{rate}
\end{equation}
where $P_{k,j}$ is the received signal power at the $k$th MT from the $j$th BS, $W$ is the channel bandwidth, $N_o$ denotes the noise power spectral density, and $I_k$ is the received interference at the $k$th MT. 
Under directional transmissions/receptions, the received signal power at the $k$th MT (UE/STA) from the $j$th BS (gNB/AP), $P_{k,j}$, is given by:
\vspace{-0.1cm}
\begin{equation}
P_{k,j}=P_{j}G_{k,j}L_{k,j}, \quad L_{k,j}= \Big(\frac{c}{4\pi f_c}\Big)^2\frac{1}{(d_{k,j})^\alpha} \label{receivedpow},
\end{equation}
where $P_{j}$ denotes the transmit power of the $j$th BS and $G_{k,j}$ is the total beamforming gain between the $j$th BS and the $k$th MT (including transmit and receive beamforming gains). $L_{k,j}$  in \eqref{receivedpow} denotes the pathloss between the $j$th BS and the $k$th MT, which depends on their distance $d_{k,j}$, the path loss exponent $\alpha$, the carrier frequency ($f_c{=}60$ GHz), and the speed of light $c$. Notice that those BSs that do not access the channel, will have a transmit power of $P_j{=}0$ (and so $P_{k,j}{=}0, \forall k$).

The total beamforming gain $G_{k,j}$ in \eqref{receivedpow} is computed by:
\begin{equation}
G_{k,j}=|\textbf{r}_{k}^H\textbf{H}_{k,j}\textbf{w}_{j}|^2, \label{G}
\end{equation}
being $\textbf{H}_{k,j}{\in} \mathbb{C}^{N\times M}$ the channel matrix that includes the complex channel gains between the antennas at the $j$th BS and the antennas at the $k$th MT. The channel matrix is given by \cite{andrews:17}:
\vspace{-0.1cm}
\begin{equation}
\textbf{H}_{k,j}=\sqrt{MNL^{-1}}\sum\nolimits_{l=1}^L \beta_l  \textbf{e}_{\text{rx},k}(\phi_{\text{rx},k}^l) \textbf{e}^H_{\text{tx},j}(\phi_{\text{tx},j}^l), \label{H}
\end{equation}
where $\beta_l$ is the complex gain of the $l$th path, $\phi_{\text{tx},j}^l$ and $\phi_{\text{rx},k}^l {\in} [0,2\pi]$ are uniformly distributed random variables that represent the angles of departure and angles of arrival, respectively, of the $l$th path, and $L$ is the total number of paths. Assuming uniform linear arrays with antenna elements spaced $\frac{\lambda}{2}$, being $\lambda{=}\frac{c}{f_c}$ the wavelength, the transmit and receive antenna array responses $\textbf{e}_{\text{tx},j}(\phi_{\text{tx},j}^l) {\in} \mathbb{C}^{M\times 1}$ and $\textbf{e}_{\text{rx},k}(\phi_{\text{rx},k}^l){\in} \mathbb{C}^{N\times 1}$ are given by \cite{andrews:17}:
\vspace{-0.1cm}
\begin{eqnarray}
&&\textbf{e}_{\text{tx},j}(\phi_{\text{tx},j}^l)= \frac{1}{\sqrt{M}}\left[ \begin{array}{cc}
1 \\ e^{-j\pi\sin(\phi_{\text{tx},j}^l)} \\ \vdots \\  e^{-j\pi(M-1)\sin(\phi_{\text{tx},j}^l)} \end{array} \right] , \\
&&\textbf{e}_{\text{rx},k}(\phi_{\text{rx},k}^l)=\frac{1}{\sqrt{N}}\left[ \begin{array}{cc}
1 \\ e^{-j\pi\sin(\phi_{\text{rx},k}^l)} \\ \vdots \\  e^{-j\pi(N-1)\sin(\phi_{\text{rx},k}^l)} \end{array} \right].
\end{eqnarray}

When the channel is dominated by a line-of-sight (LOS) component, or when the number of scatterers is small, as it occurs in mmWave bands, it becomes reasonable to design the Tx and Rx beams ($\textbf{w}_{k}$ and $\textbf{r}_{k}$) to maximize the total beamforming gain in a certain desired direction. This process is called beam-steering and, based on that, the Tx/Rx beams are given by $\textbf{w}_{j}{=}\textbf{e}_{\text{tx},j}(\phi_{\text{tx},j}^1)$, $\textbf{r}_{k}{=}\textbf{e}_{\text{rx},k}(\phi_{\text{rx},k}^1)$. In case omnidirectional reception is used, e.g., at a STA, then we denote by $\textbf{r}_{k}{=}\textbf{r}_{\text{omni}}$ the omnidirectional Rx beam.

\section{Listen Before Receive}
\label{sec:LBR}
The key idea of LBR is that the MT (e.g., the UE) is the one that manages the access to the channel and determines if the BS (e.g., the gNB) should transmit or not. However, as LBT is a regulatory requirement for operation in the unlicensed spectrum, in some regions, like Europe \cite{ETSI302567}, we define the procedure for LBR assuming that it complements LBT. Basically, LBR includes an additional physical carrier sense that is performed at the receiver side, and which can be implemented either directionally (dirLBR) or omnidirectionally (omniLBR), as for the physical carrier sense of LBT at the transmitter side. 

\begin{figure}[!t]
	\centering
	\includegraphics[width=0.45\textwidth]{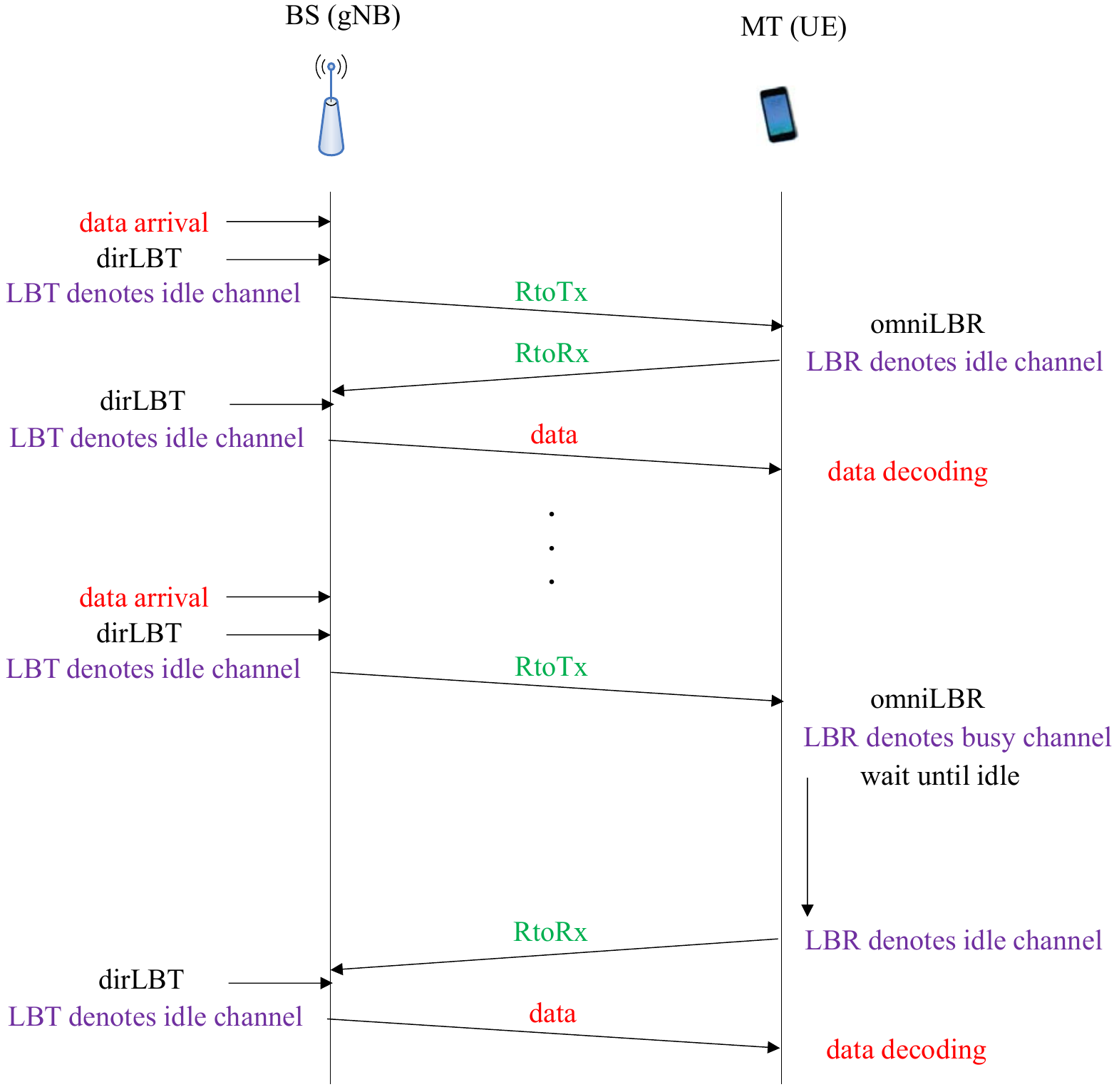}
	\caption{Call flow for omniLBR to complement dirLBT.}
	\label{fig_callflow}
	\vspace{-0.4cm}
\end{figure}

The procedure for LBR to complement LBT is as follows. Once data is ready at the BS, the BS communicates so to the MT through a "ready-to-transmit" (RtoTx) message (sent in directional manner) after a successful LBT. Then, the MT, on reception of the RtoTx message, proceeds to sense the channel with either an omnidirectional receive pattern if omniLBR (i.e., $\textbf{r}_{k}{=}\textbf{r}_{\text{omni}}$) or a directional one if dirLBR (i.e., $\textbf{r}_{k}{=}\textbf{e}_{\text{rx},k}(\phi_{\text{rx},k}^1)$), and compares the received power $S_{k}$ with a predefined threshold $P_{\text{th}}$. Note that omniLBR allows detecting nearby ongoing transmissions of other RATs or BSs, while dirLBR basically senses its receive direction. The idle/busy channel determination at the MT is as follows. If $S_{k}{\le} P_{\text{th}}$ (i.e., LBR denotes idle channel) then transmission is allowed, and a "ready-to-receive" (RtoRx) message is sent (in directional manner) to the BS. Otherwise, if $S_{k}{>} P_{\text{th}}$ (i.e., LBR denotes busy channel), transmission of the RtoRx message is postponed until the channel is sensed idle, hence deferring the data transmission. Finally, once the RtoRx message is received at the BS, and LBT is successful, data transmission can be initiated. 
Fig. \ref{fig_callflow} shows a call flow of the proposed procedure for dirLBR, when complementing dirLBT. For the first data arrival, dirLBR denotes idle channel, and data transmission is attempted after the messages exchange. For the second data arrival, dirLBR senses the channel as busy, and so transmission is deferred.

A key parameter of LBR that needs to be properly designed, is the LBR threshold $P_{\text{th}}$. If it is too high, then all the transmissions would be allowed (as far as LBT enables them). In contrast, if it is too low, all transmissions would be deferred. So, its design should be carefully analyzed depending on the deployment scenario, propagation conditions, and system parameters. Also, it could be adjusted adaptively based on reports from the BS or collision indicators (e.g., NACKs).

As both the physical carrier sensing at the BS (LBT) and the MT (LBR) can be implemented either directionally or omnidirectionally, many combinations are available for implementation: dirLBT-dirLBR, dirLBT-omniLBR, omniLBT-dirLBR, and omniLBT-omniLBR. We evaluate and compare all of them in the simulation results section.

The proposed LBR to complement LBT is similar to but differs from the directional RTS/CTS message exchange procedure that has been defined (as optional) by IEEE 802.11ad/ay \cite{wigig}, and which is common in IEEE 802.11 standards. In WiGig, RTS/CTS messages are sent directionally, but the physical carrier sense is performed omnidirectionally at all the nodes (i.e., omniLBT). 
In LBR to complement LBT, the carrier sense at the BS and the MT can be performed either directionally to avoid overprotection or omnidirectionally to detect nearby transmissions and avoid interfering them. Furthermore, as we will see in Section \ref{sec:results}, the best combination seems to be dirLBT-dirLBR, which significantly outperforms omniLBT-omniLBR. In addition to the physical carrier sense, RTS/CTS procedure also uses a virtual carrier sense, through a Network Allocation Vector (NAV), in which 
the nodes other than source or destination that hear either RTS or CTS would set their NAV according to the duration specified in RTS/CTS frames so that they would not try sensing and try to transmit anything during that period.
In LBR, we only focus on the physical carrier sense, without additional virtual carrier sensing mechanisms, because the virtual carrier sense of RTS/CTS is only useful to avoid intra-RAT interference problems.
All in all, the main objective of LBR is to act as a spectrum sharing technique for coexistence of BSs, belonging to the same or different RATs, that helps in promoting a fair behavior at the network-level, instead of a selfish one. That is, BSs are not allowed to transmit if the MT detects that the transmission could disturb nearby ongoing transmissions. Differently, the RTS/CTS mechanism was designed to control channel access and act as a way to obtain a clear channel in IEEE 802.11 random access-based networks.
RTS/CTS may be useful in Wi-Fi and WiGig networks, due to the asynchronous operational mode of the different nodes. However, in NR, nodes may be coordinated, and so maybe there is no need to add a virtual carrier sense that does only solve intra-RAT interference (NR-NR) problems.

\section{Implementation of LBR in NR}
\label{sec:impl}
In this section we first provide a brief summary of the main features of NR, to introduce basic nomenclature and concepts required to understand what follows. Afterwards, we present a potential implementation of LBR for NR-based access to unlicensed mmWave bands.

\subsection{New Radio}
With flexibility in mind, NR includes multiple numerologies to support different use cases, carrier frequencies, and deployment options. A numerology is defined by a subcarrier spacing (SCS) and a cyclic prefix (CP) overhead \cite{TR38912,TS38211}. The SCS is $15{\times}2^\mu$ kHz, and SCS from 15 kHz to 240 kHz are at least to be supported. For carrier frequencies above 40 GHz, SCS${\ge}120$ kHz is recommended \cite{zaidi:16}. The number of subcarriers per physical resource block (PRB) is fixed to 12, and the maximum number of PRBs is 275 in NR Phase 1 (Release 15).
On the other hand, the frame length is set to 10 ms, and a frame is composed of 10 subframes of 1 ms each, to maintain backward compatibility with LTE. The main difference concerning LTE is that, according to the different SCS, the OFDM symbol length varies for every numerology (see Table \ref{fig_num}), which allows reducing the transmission time duration. A slot is defined as 14 OFDM symbols, so the slot length varies with the numerology as well. Table \ref{fig_num} shows the configuration of different numerologies. $\mu{=}0$ corresponds to the LTE system configuration, while $\mu{>}0$ enables larger bandwidth and shorter slot length, which is useful for mmWave bands as well as for operation in the unlicensed spectrum. 

\begin{table} [!t]
	\renewcommand{\arraystretch}{1.1}
	\vspace{0.2cm}
	\caption{\textsc{Numerologies for NR.}}
	\vspace{-0.2cm}
	\label{fig_num}
	\scriptsize
	\begin{center}		
		\begin{tabular}{p{3.5cm} || p{0.45cm}  | p{0.45cm} | p{0.45cm} | p{0.45cm} | p{0.45cm}  }
			\textbf & $\mu$=0 & $\mu$=1 &$\mu$=2 &$\mu$=3 &$\mu$=4  \\
			\hline
			\hline
			subcarrier spacing [kHz] & 15 & 30  & 60 & 120 & 240  \\
			\hline
			OFDM symbol length [us] & 66.67 & 33.33 & 16.67 & 8.33 & 4.17 \\
			\hline
			cyclic prefix [us] & $\backsim$4.8 & $\backsim$2.4 & $\backsim$1.2  & $\backsim$0.6 & $\backsim$0.3 \\
			\hline
			frame length [ms] & 10  & 10  & 10 & 10  & 10 \\
			\hline
			number of subframes in a frame & 10 & 10 & 10 & 10 & 10 \\
			\hline
			number of slots in a subframe & 1 & 2& 4& 8 & 16 \\
			\hline
			slot length [us] & 1000 & 500& 250 &125 & 62.5\\
			\hline
			number of OFDM symbols in a slot & 14 & 14 & 14 & 14 & 14  \\
			\hline
			number of subcarriers in a PRB & 12 & 12 & 12 & 12 & 12  \\
			\hline
			PRB width [MHz] & 0.18 & 0.36 & 0.72 & 1.44 & 2.88  \\
			\hline
		\end{tabular}
		\vspace{-0.2cm}
	\end{center}
\end{table}

Another interesting feature that has potential for NR-based access to unlicensed spectrum bands, is the self-contained slot \cite{TS38211}. The self contained slot may include DL control, data (DL and/or UL), and UL control, within the same slot, as shown in Fig. \ref{fig_slotheaders}. Therefore, for example, it allows DL data and the corresponding ACK/NACK (in UL control part) to be sent in the same slot. Also, it enables the UE to receive the UL grant (in DL control part) and transmit UL data in the same slot. 

To reduce the delays associated to the message exchange in Fig. \ref{fig_callflow}, we propose to implement LBR in conjunction with the self-contained slot for efficient NR-based access to unlicensed spectrum, as detailed next.

\subsection{LBR Implementation based on Self-Contained Slot}
\begin{figure}[!t]
	\centering
	\includegraphics[width=0.42\textwidth]{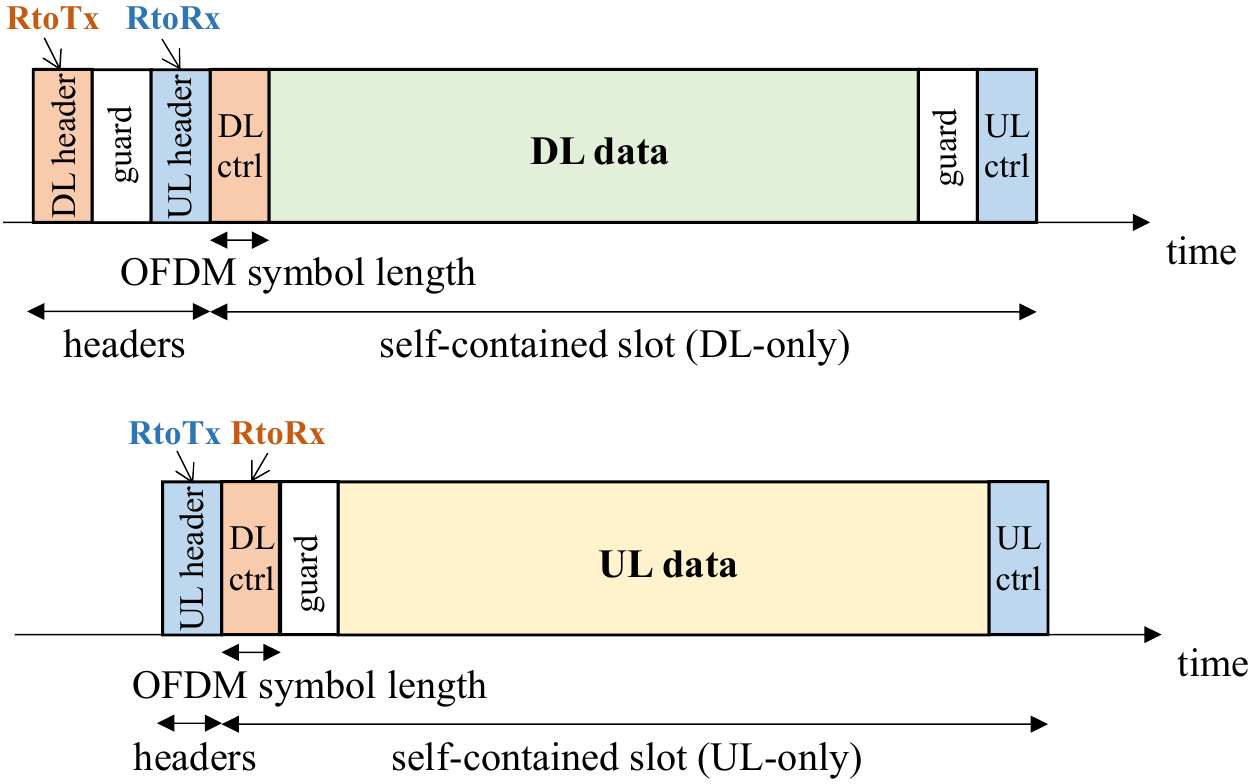}
	\caption{LBR to complement LBT with the self-contained slot for DL-only (top) and UL-only (bottom) slots.}
	\label{fig_slotheaders}
	\vspace{-0.3cm}
\end{figure}

LBR to complement LBT could be easily implemented in case that, before the self-contained slot structure, DL/UL headers were included in a preparation stage, as suggested by Qualcomm in \cite{qualcomm:17}, and as shown in Fig. \ref{fig_slotheaders}. For the DL access, the RtoTx/RtoRx messages would be included in the DL/UL headers, respectively, without incurring additional delays. For UL access, only the UL header would be required to send the RtoTx, and the RtoRx message could be directly sent in the DL control part.

In case that no headers were available, LBR to complement LBT could be implemented using the DL/UL control parts. 
For DL access, the RtoTx message would be sent within the DL control part in slot $n$, and the RtoRx message would be sent within the UL control part of the same slot $n$. Then, once the RtoRx message is received at the gNB, DL data could be transmitted within slot $n{+}1$. This way, only one slot would be left empty from the point in which the data is ready at the gNB and it is actually transmitted over the air. Note, indeed, that the slot lengths are reduced as the SCS increases (see Table \ref{fig_num}). Furthermore, there is no need to use reservation signals, because the gNB has to wait for the RtoRx message and, once it is received, channel access can be attempted immediately within the subsequent slot. 

For UL access, the RtoTx message already follows the UL scheduling request (SR) in LTE and NR\footnote{In LTE, UL transmissions follow a two-way handshake message exchange that consists of a SR sent by the UE and an UL grant sent by the eNB, before the scheduled UL data transmission can go through.}. Similarly to the case of having an UL header, SR and RtoTx would be sent in the UL control part of slot $n$. Then, the RtoRx message could be sent jointly with the UL grant within the DL control part in slot $n{+}1$, followed by UL data. Therefore, for UL access, LBR does not represent any additional inefficiency or complexity, since it only requires the gNB to sense the channel before sending the UL grant to UE.

To compute how much percentage of time is left unoccupied in the DL access because the gNB is waiting for the RtoRx message from the UE, when no DL/UL headers are used, we need to consider the maximum channel occupancy time (MCOT) for unlicensed spectrum access and the different slot lengths in NR. 
The MCOT takes a value of 9 ms in the 60 GHz band \cite{ETSI302567}. Therefore, the percentage of unoccupied time is of 1.38$\%$ for $\mu{=}3$ (SCS=120 kHz) and 0.69$\%$ for $\mu{=}4$ (SCS=240 kHz). Accordingly, the implementation of LBR in conjunction with the self-contained slot, either including headers or not, allows for efficient access to the medium.

\vspace{0.1cm}
\section{Simulation Results}
\label{sec:results}
To illustrate the benefits of the proposed LBR, we consider a dense indoor network deployment, composed of $K$ BS-MT pairs that are randomly deployed in a $10{\times}10$ $\text{m}^2$ area \cite{cai:10}, in which the BS-MT pair distance is $d_{k,k}{=}4$ m, $\forall k$. Performance of the DL transmission is evaluated, assuming LOS between the different BS-MT pairs, and that BSs operate at $f_c{=}60$ GHz with $W{=}1$ GHz bandwidth and an available power of $P_j{=}10$ dBm, $\forall j$. The pathloss model in \eqref{receivedpow} with $\alpha{=}2$ is adopted, which resembles IEEE 802.11ad pathloss model \cite{wigig}. The noise power spectral density is $N_o{=}{-}174$ dBm/Hz. 

To simplify the beamforming gain computation (i.e., $G_{k,j}$ in \eqref{G}), we characterize the Tx and Rx beams through a widely-used and simple model: the cone plus circle model in a two-dimensional scenario \cite{cai:10}. Based on that, the directional antenna pattern of every $j$th BS consists of a mainlobe with beamwidth $\theta_\text{tx}^j$ and gain 
$G_{\text{tx},\text{m}}^j$ and a sidelobe of beamwidth $2\pi{-}\theta_\text{tx}^j$ and gain $G_{\text{tx},\text{s}}^j$ (see \cite[Sect. IV.C]{andrews:17}). Similarly, the antenna pattern at every $k$th MT is characterized by the beamwidth and gain of the mainlobe, $\theta_\text{rx}^k$ and $G_{\text{rx},\text{m}}^k$, and the beamwidth and gain of the sidelobe, $2\pi{-}\theta_\text{rx}^k$ and $G_{\text{rx},\text{s}}^k$.
For every link, depending on the orientation of the mainlobes at the $j$th BS and the $k$th MT, either $G_{\text{tx},\text{m}}^j$ or $G_{\text{tx},\text{s}}^j$ (that conforms the $j$th BS Tx beam gain in the direction of the $k$th MT, $G_{\text{tx}}^{j,k}$), and either $G_{\text{rx},\text{m}}^k$ or $G_{\text{rx},\text{s}}^k$ (that come into $G_{\text{rx}}^{k,j}$, i.e., the $k$th MT Rx beam gain in the $j$th BS direction), are used to compute the total beamforming gain in \eqref{G} as $G_{k,j}{=}\beta_1G_{\text{rx}}^{k,j}G_{\text{tx}}^{j,k}$, being $\beta_1$ the complex gain of the LOS path (see \eqref{H}). 

For simulations, we fix the Tx and Rx beam gains to $G_\text{tx,m}^j{=}10$ dB and $G_\text{rx,m}^k{=}10$ dB, respectively, $\forall j, k$, and ideal antenna radiation efficiency is assumed (i.e., $G_\text{tx,s}^j{=}0$, $G_\text{rx,s}^k{=}0$). The number of BS-MT pairs ($K$) and their Tx and Rx mainlobe beamwidths ($\theta_\text{tx}^j$ and $\theta_\text{rx}^k$) are varied through simulations. 

As baseline methods to benchmark, omniLBT and dirLBT techniques are considered. Then, we evaluate all the possible combinations of LBT and LBR: omniLBT-omniLBR, omniLBT-dirLBR, dirLBT-omniLBR, dirLBT-dirLBR, as indicated in the legends.
The ED threshold for LBT and LBR, normalized by the maximum antenna gain\footnote{Directional transmissions are considered, but the carrier sense can be done either directional or omnidirectional. Thus, a normalized ED threshold of ${-74}$ dBm, by taking into account the array gain used for sensing, corresponds to ${-74}$ dBm for omniLBT/omniLBR and to ${-64}$ dBm for dirLBT/dirLBR.}, is set to ${-74}$ dBm. We do not emulate backoff processes, and simply consider how many BS-MT pairs can reuse the spectrum according to the different channel access procedures. Simulation results are averaged among 1000 random deployments, 
with random start time. 
As performance metrics we use the sum-rate (i.e., the sum of data rates: $\sum_{\forall k} {R_k}$, being $R_k$ defined in \eqref{rate}, which measures how many pairs can simultaneously access the channel and their data rate) and the mean-rate during channel access (i.e., the average of those $R_k$ such that $R_k{>}0$, which illustrates the quality-of-service (QoS) obtained by the pairs that get access to the channel).

\begin{figure}[!t]
	\centering
	\includegraphics[width=0.39\textwidth]{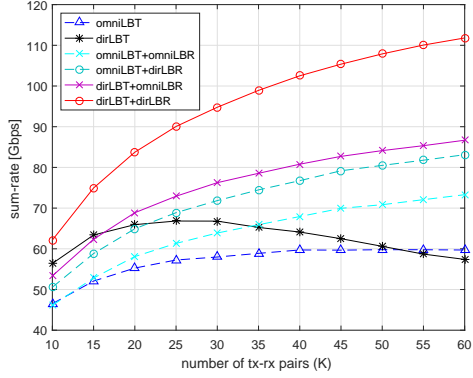}
	\caption{Sum-rate (in Gbits/s) vs. the number of BS-MT pairs ($K$). }
	\label{fig1}
	\vspace{-0.45cm}
\end{figure}

\begin{figure}[!t]
	\centering
	\includegraphics[width=0.39\textwidth]{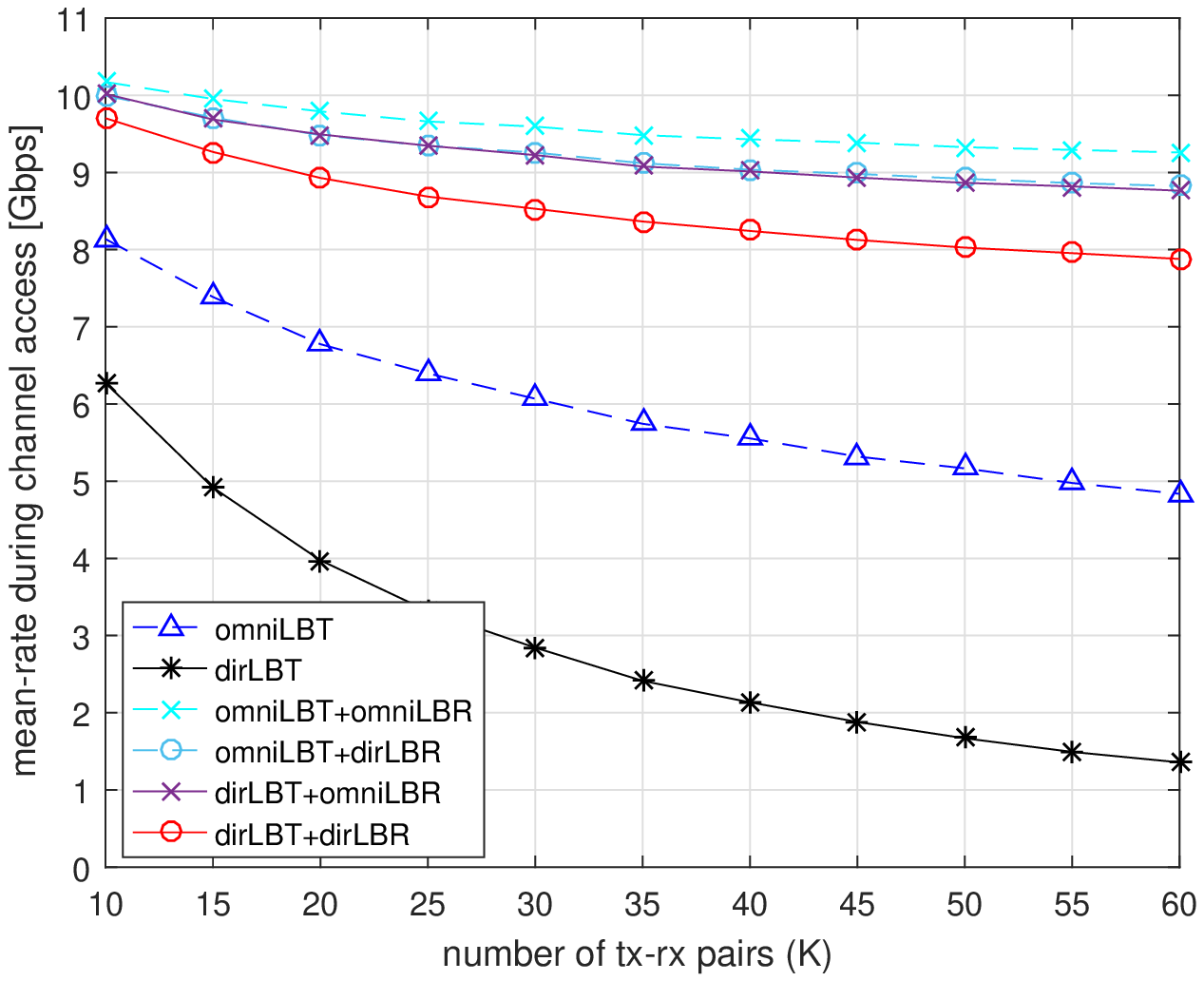}
	\caption{Mean-rate (in Gbits/s) vs. the number of BS-MT pairs ($K$). }
	\label{fig2}
	\vspace{-0.45cm}
\end{figure}

Fig. \ref{fig1} and Fig. \ref{fig2} show the sum-rate and the mean-rate during channel access, respectively, versus $K$, for $\theta_\text{tx}^j{=}60^o$ and $\theta_\text{rx}^k{=}90^o$, $\forall j, k$. 
For low $K$, omniLBT achieves the lowest performance due to its overprotective behavior that prevents access of a large number of pairs. DirLBT allows for an improved spatial reuse, and hence an enhanced sum-rate and mean-rate for low $K$. However, as $K$ increases, the sum-rate of dirLBT decreases due to an excess of hidden node problems. Notably, all the procedures that use LBR allow for a significantly improved sum-rate and mean-rate as compared to dirLBT and omniLBT when $K$ increases. Results show that the impact of hidden nodes, which appeared with dirLBT for high $K$, is properly addressed by LBR and the performance is not penalized by the network density. This confirms that the receiver has useful information that needs to be properly exploited for coexistence in unlicensed mmWave bands. Among the LBT-LBR combinations, dirLBT-dirLBR is shown to get the best sum-rate for all the emulated network densities, since it is the one that allows large spatial reuse while avoiding hidden node problems. In terms of mean-rate, the LBT-LBR solutions that include omnidirectional sensing at some of the nodes achieve a slightly better performance.

\begin{figure}[!t]
	\centering
	\includegraphics[width=0.39\textwidth]{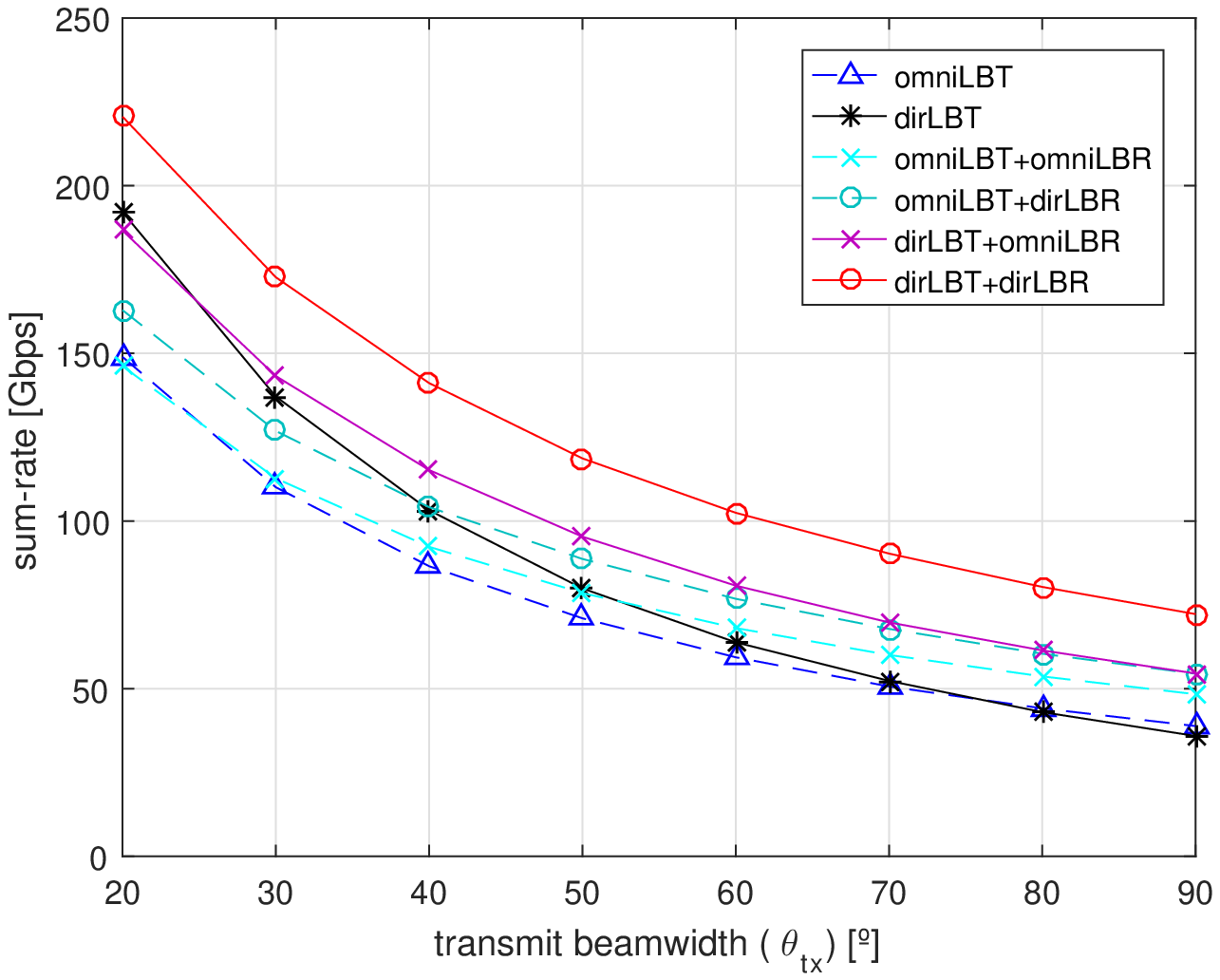}
	\caption{Sum-rate (in Gbits/s) vs. Tx beamwidth ($\theta_\text{tx}^j$) for $K{=}40$.}
		\label{fig3}
	\vspace{-0.45cm}
\end{figure}

\begin{figure}[!t]
	\centering
	\includegraphics[width=0.39\textwidth]{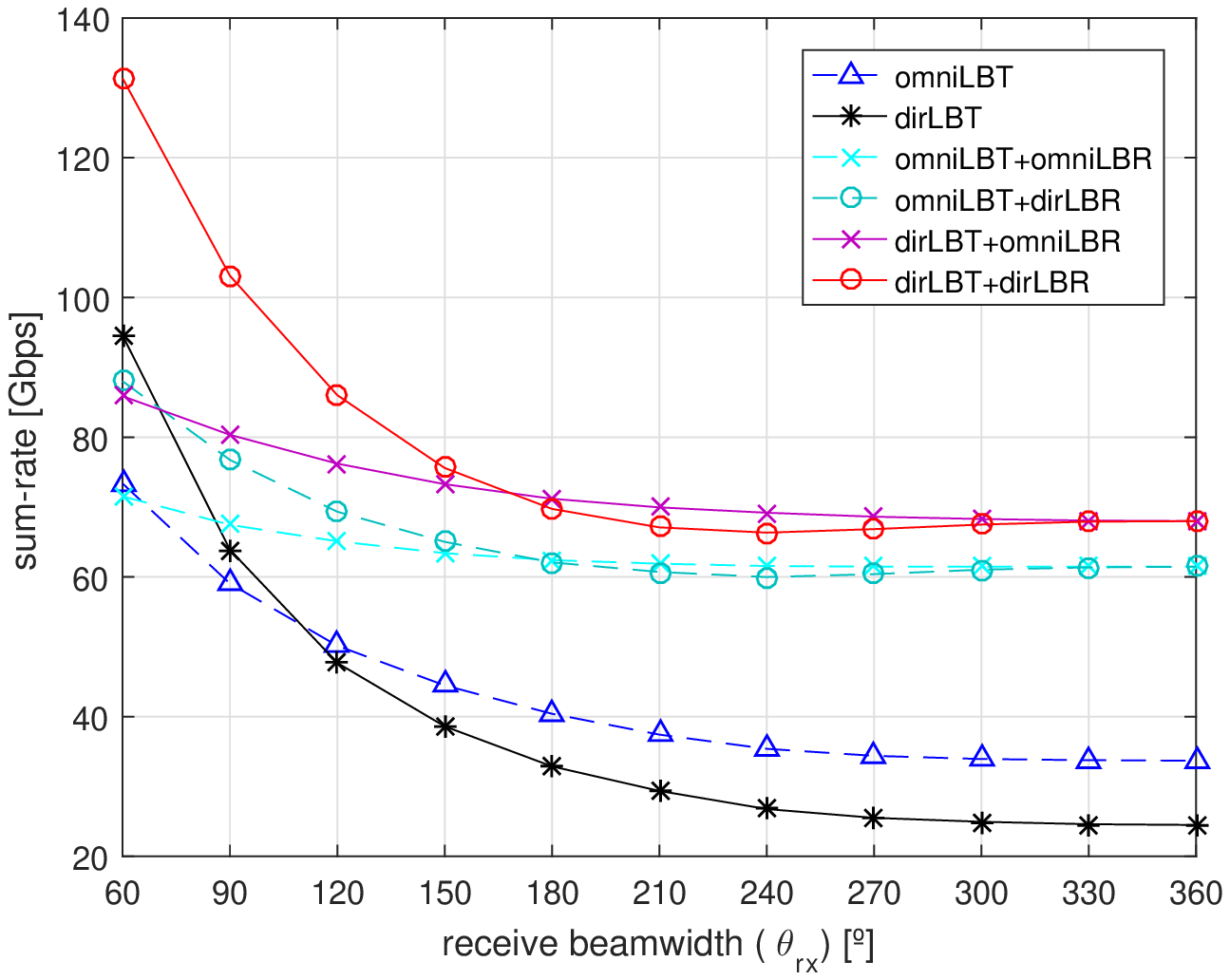}
	\caption{Sum-rate (in Gbits/s) vs. Rx beamwidth ($\theta_\text{rx}^k$) for $K{=}40$.}
	\label{fig4}
	\vspace{-0.6cm}
\end{figure}

Finally, we vary the Tx and Rx beamwidths. Fig. \ref{fig3} displays the sum-rate versus the Tx beamwidth $\theta_\text{tx}^j$ (in $^o$), for $K{=}40$ and $\theta_\text{rx}^k{=}90^o$, $\forall k$. Fig. \ref{fig4} depicts the sum-rate versus the Rx beamwidth $\theta_\text{rx}^k$ (in $^o$), for $K{=}40$ and $\theta_\text{tx}^j{=}60^o$, $\forall j$. Again, we observe that dirLBT-dirLBR provides the largest sum-rate performance for different Tx/Rx beamwidths. When omnidirectional reception is considered ($\theta_\text{rx}^k{=}360^o$), dirLBR and omniLBR performances converge.

\section{Conclusions}
\label{sec:conclusions}
This paper proposes an LBR technique for channel access to unlicensed/shared spectrum bands with directional transmissions and receptions. The key idea of LBR is that the receiver can either allow or defer the transmission based on its carrier sense. 
We propose to implement LBR in conjunction with LBT and the self-contained slot in 5G NR, which is shown to enable efficient access to the shared channel. Also, we evaluate multiple combinations of LBT and LBR, and conclude that dirLBT-dirLBR achieves the best performance.

\section{Acknowledgments}
\label{sec:ack}
\small
The research leading to these results was supported by InterDigital, Inc. Also, it was partially funded by Spanish MINECO grant TEC2017-88373-R (5G-REFINE) and Generalitat de Catalunya grant 2017 SGR 1195.

\bibliography{references}
\bibliographystyle{ieeetran}

\end{document}